\documentclass[journal=jacsat,manuscript=article]{achemso}

\usepackage[version=3]{mhchem}

\author{Ming Li}
\affiliation{Key Laboratory of Quantum Information, University of Science and Technology of China, CAS, Hefei, 230026, People's Republic of China}
\alsoaffiliation{Synergetic Innovation Center of Quantum Information $\&$ Quantum
Physics, University of Science and Technology of China, Hefei, Anhui
230026, China}
\author{Xiao Xiong}
\affiliation{Key Laboratory of Quantum Information, University of Science and Technology of China, CAS, Hefei, 230026, People's Republic of China}
\alsoaffiliation{Synergetic Innovation Center of Quantum Information $\&$ Quantum
Physics, University of Science and Technology of China, Hefei, Anhui
230026, China}
\author{Le Yu}
\affiliation{Key Laboratory of Quantum Information, University of Science and Technology of China, CAS, Hefei, 230026, People's Republic of China}
\alsoaffiliation{Synergetic Innovation Center of Quantum Information $\&$ Quantum
Physics, University of Science and Technology of China, Hefei, Anhui
230026, China}
\author{Chang-Ling Zou}
\affiliation{Key Laboratory of Quantum Information, University of Science and Technology of China, CAS, Hefei, 230026, People's Republic of China}
\alsoaffiliation{Synergetic Innovation Center of Quantum Information $\&$ Quantum
Physics, University of Science and Technology of China, Hefei, Anhui
230026, China}
\author{Yang Chen}
\affiliation{Key Laboratory of Quantum Information, University of Science and Technology of China, CAS, Hefei, 230026, People's Republic of China}
\alsoaffiliation{Synergetic Innovation Center of Quantum Information $\&$ Quantum
Physics, University of Science and Technology of China, Hefei, Anhui
230026, China}
\author{Di Liu}
\affiliation{Key Laboratory of Quantum Information, University of Science and Technology of China, CAS, Hefei, 230026, People's Republic of China}
\alsoaffiliation{Synergetic Innovation Center of Quantum Information $\&$ Quantum
Physics, University of Science and Technology of China, Hefei, Anhui
230026, China}
\author{Lan-Tian Feng}
\affiliation{Key Laboratory of Quantum Information, University of Science and Technology of China, CAS, Hefei, 230026, People's Republic of China}
\alsoaffiliation{Synergetic Innovation Center of Quantum Information $\&$ Quantum
Physics, University of Science and Technology of China, Hefei, Anhui
230026, China}
\author{Guo-Ping Guo}
\affiliation{Key Laboratory of Quantum Information, University of Science and Technology of China, CAS, Hefei, 230026, People's Republic of China}
\alsoaffiliation{Synergetic Innovation Center of Quantum Information $\&$ Quantum
Physics, University of Science and Technology of China, Hefei, Anhui
230026, China}
\author{Xi-Feng Ren}
\affiliation{Key Laboratory of Quantum Information, University of Science and Technology of China, CAS, Hefei, 230026, People's Republic of China}
\alsoaffiliation{Synergetic Innovation Center of Quantum Information $\&$ Quantum
Physics, University of Science and Technology of China, Hefei, Anhui
230026, China}
\email{renxf@ustc.edu.cn}
\author{Guang-Can Guo}
\affiliation{Key Laboratory of Quantum Information, University of Science and Technology of China, CAS, Hefei, 230026, People's Republic of China}
\alsoaffiliation{Synergetic Innovation Center of Quantum Information $\&$ Quantum
Physics, University of Science and Technology of China, Hefei, Anhui
230026, China}

\title{Near-field collection of fluorescence of quantum dots with a fiber-integrated multimode plasmonic probe }

\makeatother

\begin{document}

\newpage{}

\begin{abstract}
 Strong light-matter interaction and high-efficiency optical collection
 of fluorescence from quantum emitters are crucial topics in quantum and nanophotonic
 fields. High-quality cavities, dispersive photonic crystal waveguides
 and even plasmonic structures have been used to enhance the interaction
 with quantum emitters, thus realize efficient collection of the
 fluorescence. In this work, a new method is proposed to collect the
 fluorescence of quantum dots (QDs) with a fiber-integrated multimode
 silver nanowire (AgNW) waveguide. Fluorescence lifetime measurement
 is performed to investigate the coupling between QDs and different
 plasmonic modes. Compared with far-field collection method, the AgNW-fiber probe
 can realize near-unity collection efficiency theoretically. This fiber-integrated plasmonic
 probe may be useful in the area of nanophotonics and also promising
 for quantum information devices.
\end{abstract}
\maketitle

\section{Keywords}

Surface plasmon polaritons, Quantum dots, Silver nanowire

\section{Introduction}

In nanophotonics, single emitters including QDs \cite{Reviewqd}, dye molecules or other quantum emitters have been widely investigated in the fields of solar cells, biologic labeling\cite{biolabel}, quantum information\cite{photonsource,chenxd}, etc. Experimentalists have spared no effort to control their properties and improve the signal collection efficiency, both of which require strong light-matter interaction\cite{principleofnano-optics}. Traditionally, people study the quantum emitters by collecting the fluorescence at far-field with lens. Along with the development of integrated optics\cite{Obrien}, integrated photon source, as well as its near-field manipulation and detection, are possible by incorporating the single emitters with integrated photonic structures. More importantly, due to the strong light confinement in waveguides or cavities, the emission of quantum emitters can be efficiently collected and directly utilized for the followed processing.

Recently, the coupling between quantum emitters and dielectric waveguides have been demonstrated\cite{qd-waveguidenanoletter,qdwaveguideprl}. Schemes aimed to increase the collection efficiency by placing the emitters at the waveguide end facet have been proposed as well, for example, a theoretical calculation with fiber taper\cite{fibertipcollection}, and a movable diamond single photon source\cite{diamondsource}. However, in all these studies on which dielectric waveguide is based, the density of state is limited by the diffraction of light, resulting in the weak interaction between guided optical modes and quantum emitters. Thus, the collection efficiency is limited. Even though the weak interaction can be compensated by slowing down the group velocity of light via strong dispersive photonic crystal waveguide\cite{photonicscrystal}, the widely spread mode area (usually micrometer scale) makes the local control over quantum emitter beyond the optical diffraction limit quite difficult.

Fortunately, plasmonic waveguide has natural advantages in addressing such problems. Compared to the dielectric waveguide, plasmonic waveguide has much higher mode density, leading to stronger coupling between quantum emitters and the waveguide. And the smaller plasmonic mode area also makes the local operation of quantum emitters possible. Such strong coupling doesn't have to rely on a cavity with ultra-high quality factor, and the enhancement of surface plasmon is broadband. These advantages have promoted the development of quantum plasmonics\cite{quantumplasmonics}. For example, experimentally, the coupling of quantum emitters, including QDs, nitrogen-vacancy centers, with metallic waveguide has been investigated\cite{qdnature,nvAgNW,qe-groove,lifetimexuhongxing}. It has also been demonstrated that quantum states can be preserved by surface plasmon polariton\cite{nature2002,nanoletterstatistics,liming}. People have also studied the polarization property of the emitted photons with different plasmonic structure\cite{wangllpolarization}, the quantum statistical features of the emitted photons, and the distinguishability of single plasmons emitted by different QDs near a plasmonic waveguide\cite{Liqiang}. It's been proposed that obtaining near-unity collection efficiency is possible by coupling QD with a plasmonic tip\cite{DEchangprb}.

In this work, we develop a new method to collect the fluorescence of $ZnS$ coated colloidal $CdSe$ QDs efficiently with a fiber-integrated plasmonic probe. Among various plasmonic waveguides, AgNWs\cite{AgNWlocgicgate,xiongxiaoreview,zhedareview} have been widely implemented in nanophotonics due to its low loss, uniformity and easy preparation process. Here, we integrate the AgNW together with a fiber taper to collect the fluorescence of the QDs. The probe can transport light beyond the diffraction limit with high efficiency, which outperforms the traditional metal-coated dielectric tips.We make a detailed analysis on the efficiency of the probe, and characterize the performances of different plasmonic modes during the coupling process by measuring the lifetime of the QDs. The large decrease of the lifetime reflects the strong coupling strength between the QDs and the AgNW. Our plasmonic probe can be a promising candidate for both high-eficiency collection and optical high-resolution manipulation of quantum emitters.

\section{Experiment and Results}

\begin{figure*}[htb]
\includegraphics[width=14cm]{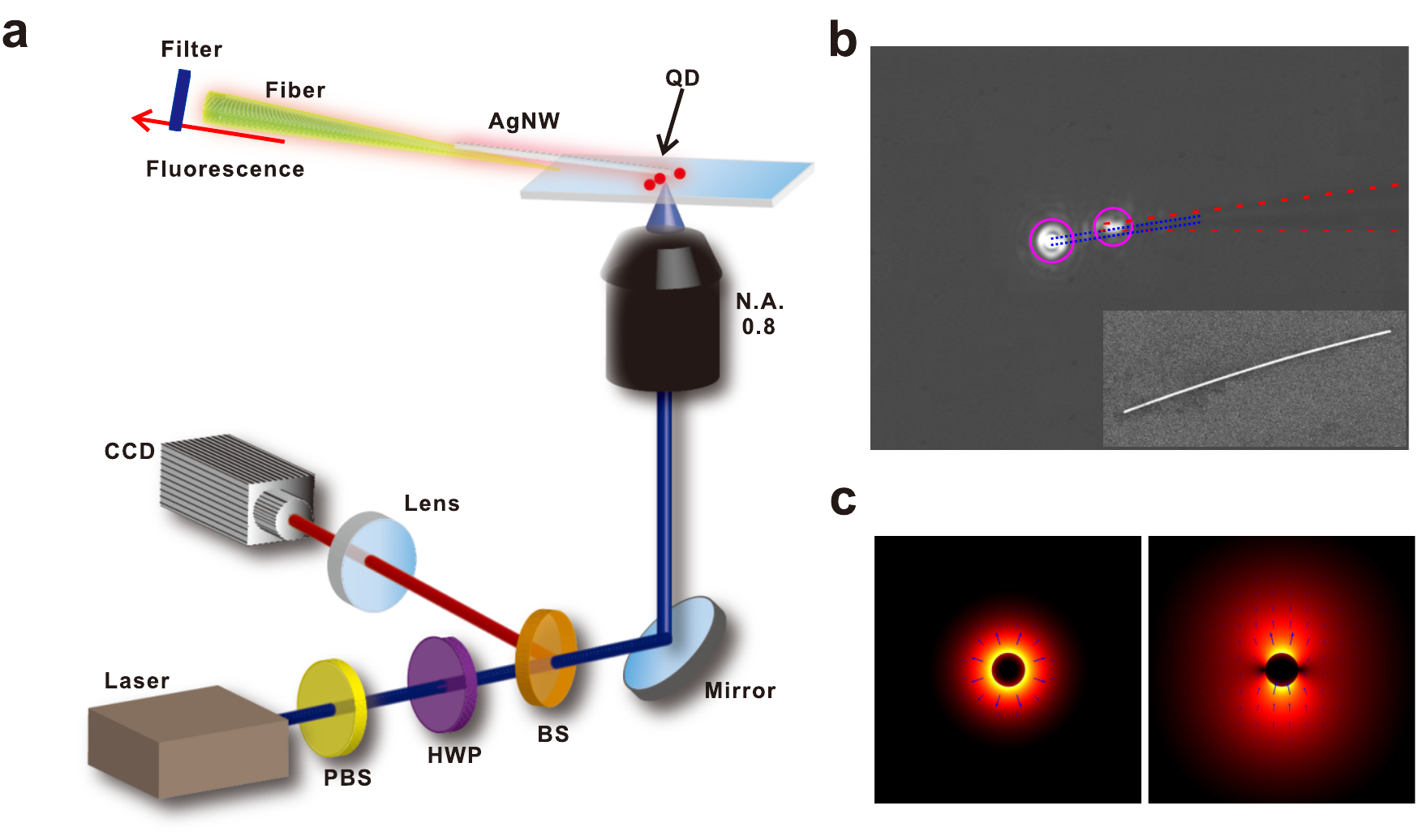} 

\caption{(color online)Experimental setup. (a) A $395nm$ ps-pulsed laser is focused by an objective lens on the substrate. PBS and HWP are used to control the polarization of the laser. The fluorescence is collected by the AgNW-fiber probe and filtered with a $561\:nm$ long pass filter to block the laser before detection. We used a single photon detector or a spectrometer to record the single photons emitted by the QDs. A multichannel analyzer is used to record the time interval between the laser pulse and the fluorescence photon, thus giving the lifetime of the fluorescence. PBS, polarization beamsplitter; HWP, half waveplate; BS, beamsplitter. (b) CCD image of the hybrid probe with laser light launched from the fiber. Magenta circles show the positions of the fiber taper end and the AgNW end. Inset: SEM (scanning electron microscope) image of a $100\:nm$-radius AgNW. (c) Eigenmodes of the AgNW. The AgNW can support three eigenmodes, $TEM_{0}$ and two degenerate modes, $TE_{1}$ and $TM_{1}$. Here, we give the electric field distributions of $TEM_{0}$ (left panel) and $TM_{1}$ (right panel).}
\end{figure*}

The experimental setup is shown in Figure 1a. The $CdSe$ QDs are spin-coated onto a silica substrate. A $395nm$ ps-pulsed laser is focused on the QD ensemble by an objective lens to excite the QDs. Unlike the previous experiment \cite{lenscollection}, which also uses objective lens to collect the fluorescence, here we use a fiber-integrated plasmonic probe to collect the signal. Our fiber-integrated plasmonic probe consists of a fiber taper and a AgNW as shown in Figure 1b. The AgNW is adhered to the tip of fiber taper with epoxy resin glue (details in the supporting information). Adiabatic coupling between the dielectric modes in the fiber taper and the plasmonic modes in AgNW ensures the highly efficient transport of light\cite{dongfibercouple,zhedafibercouple}, thus makes the detection of single emitters possible. A three-dimensional piezo-stage is used to control the distance between the hybrid probe and the excited QDs. After adjusting the probe to an appropriate position, the signal from the fiber output is analyzed with the laser being filtered by a $561nm$ long-pass filter.

First of all, we guided the collected signal into the spectrometer. For comparison, we collected the QDs emission with both objective lens ($N.A.=0.8$) and AgNW-fiber probe, and the measured spectra are shown in Figure 2. The spectra are centered at $655nm$, which is consistent with the fluorescence spectrum of $CdSe$ QD. This clearly demonstrates that we can collect the fluorescence of quantum emitters with such a probe. With a fixed excitation laser power, the counts collected by the probe and the objective lens ($N.A.=0.8$) are $3.67 \times 10^{5}/s$ and $9.75 \times 10^{5}/s,$ respectively. In consideration that the collection area of the probe is smaller than the objective lens, the collection efficiency of the plasmonic probe is at least at the same order with the objective lens.

\begin{figure}[htb]
\includegraphics[width=8cm]{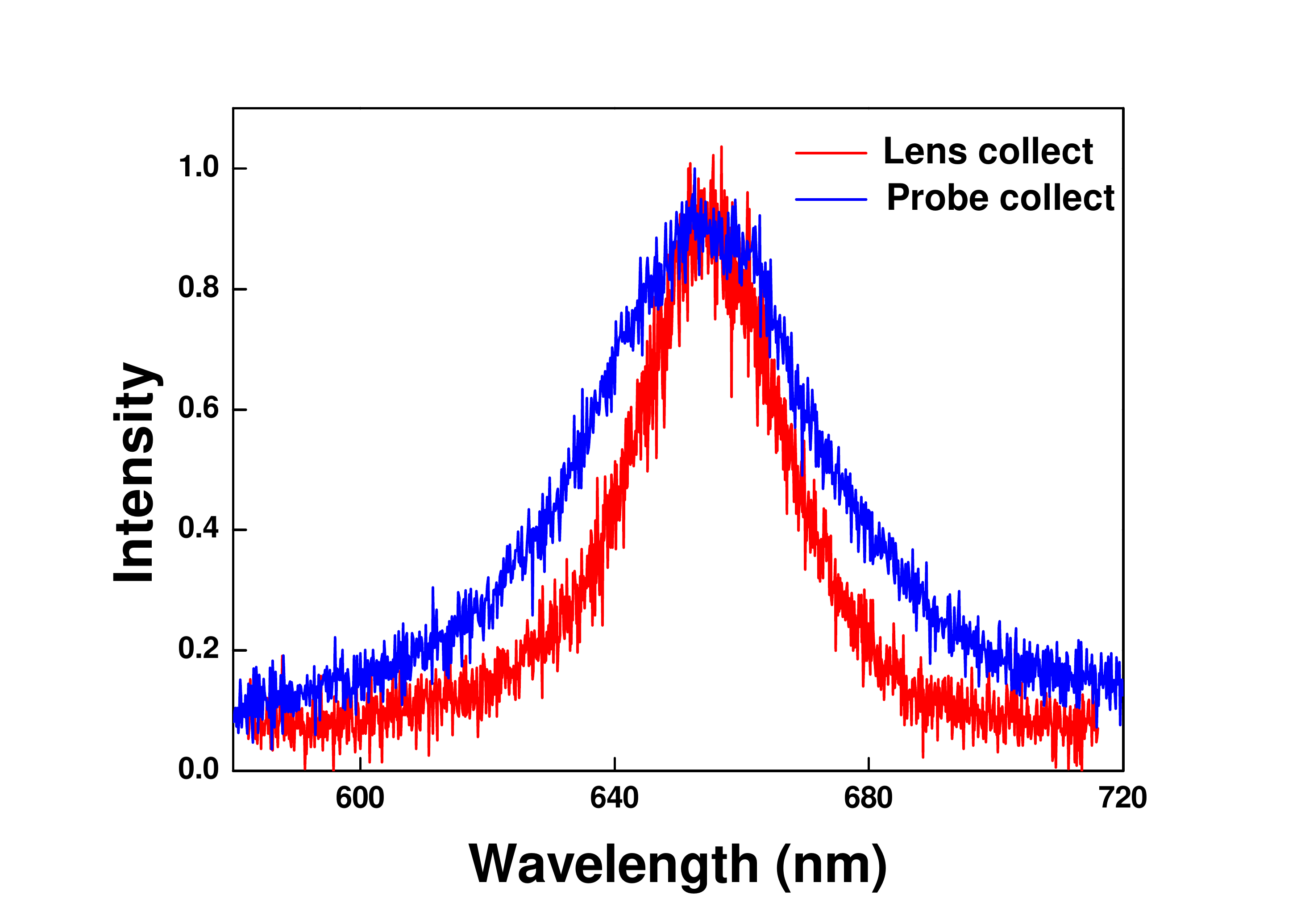}\caption{(colr online) Fluorescence spectra of CdSe quantum dot. Ren line: Fluorescence of QDs on the substrate collected by the objective lens. Blue line: Fluorescence collected by the AgNW-fiber probe. }
\end{figure}

Then, we further proved that the collected photons come from the near- field coupling between QDs and surface plasmon mode. A direct evidence is the change of the spectrum of the QD fluorescence. From Figure 2, we can see that the probe-collected spectrum is wider than that of QD on silica substrate. This mainly results from the enhanced non-radiative process of QDs coupled to the charge reservoir AgNW\cite{lifetimexuhongxing,quenching}, which caused the quenching of the QDs and the broadening of the fluorescence spectrum.

Another strong evidence for the strong near-field coupling strength is the decrease of the QD lifetime. According to the Fermi's golden rule, the spontaneous emission rate of a dipole to mode $i$ is proportional to the square of the coupling strength and the density of state. The lifetime of the dipole is defined as the inverse of the spontaneous emission rate. So the lifetime $\tau$ has a negative relation with the mode density and the coupling strength. If we put a probe near the quantum emitter, the dipole not only interacts with the continuum vacuum field in free space, but also interacts with the plasmonic modes supported by the probe. The additional channel causes the increase of the spontaneous emission rate and the decrease of the lifetime of the florescence signal. In our experiment, we measured the lifetime of the QD ensemble in different situation: (1) QDs on silica substrate, with objective lens collection. (2) QDs on silica substrate, with plasmonic probe collection. (3) QDs attached on fiber taper, with fiber taper collection.

\begin{figure*}[htb]
\includegraphics[width=16cm]{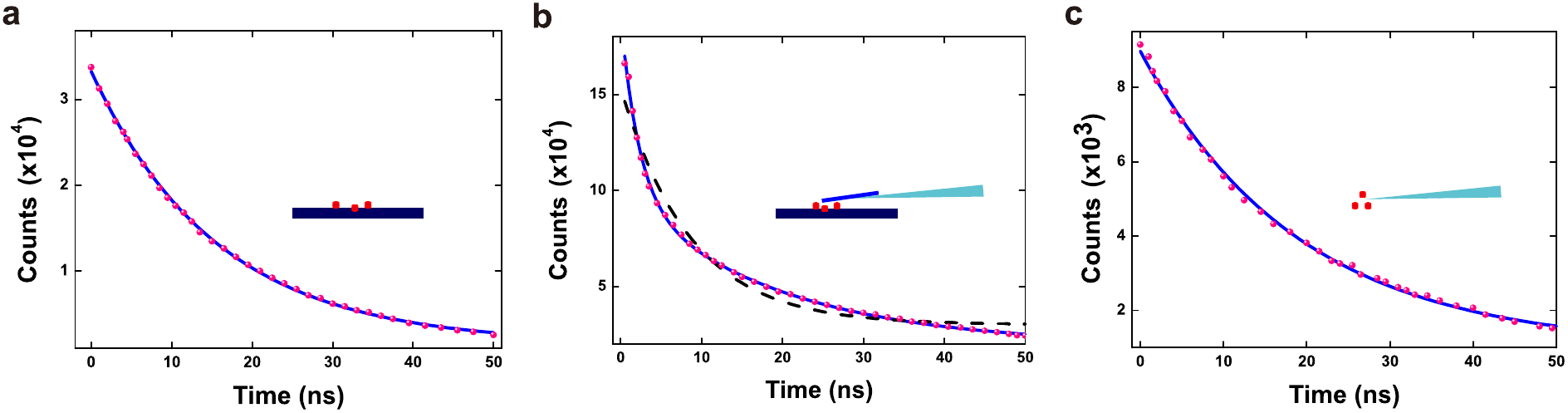}

\caption{(color online)Lifetime of the collected fluorescence. (a) QDs on silica substrate, with objective lens collection. Intensity-time relation is fitted with a single exponential function $f_{1}$. Fitted lifetime is $\tau_{0}=15.25 \pm 0.10\:ns$. (b) QDs on silica subsrate, with plasmonic probe collection. Solid line is fitted with the sum of two exponential functions $f_{2}$. $\tau_{1}=2.21 \pm 0.07\:ns$, $\tau_{2}=19.25 \pm 0.68\:ns$. Dashed line is fitted with a single exponential function $f_{1}$ with $\tau_{0}=8.61 \pm 0.35\:ns$. The large mismatch between the data and the fitted curve using $f_{1}$ implies that the collected photons come from at least two coupling processes with significant different lifetimes. (c) Intensity-time relation for fluorescence collected by fiber taper. The data is fitted with function $f_{1}$ with $\tau_{0}=19.02 \pm 0.26\:ns$. In the experiment, the data acquisition time is different for different cases.}
\end{figure*}

First we measured the averaged lifetime of $CdSe$ QD ensemble on the silica substrate. The data are perfectly fitted with a single exponential function $f_{1}=I_{0}+Ae^{-t/\tau_{0}}$, and the fitted lifetime is $15.25 \pm 0.10ns$ (See Figure 3a). Then we moved our plasmonic probe towards the excited QDs and maximize the collection efficiency. Different from the above case, the data can not be effectively fitted with a single exponential function, which results in an averaged lifetime of $8.61 \pm 0.35ns$ (see Figure 3b, black dash line). It means the collected photons are a mixture of photons with different lifetimes. So we used the sum of two exponential functions to fit the experimental data (see Figure 3b, blue solid line). The function is $f_{2}=I_{0}+A_{1}e^{-t/\tau_{1}}+A_{2}e^{-t/\tau_{2}}$, where $\tau_{1}$and $\tau_{2}$ are the lifetimes of QDs in different environments respectively, $A_{1}$ and $A_{2}$ are the weight of the collected photons from the two kind of QDs (Here, $A_{1}/A_{2}=1.17$). The fitted $\tau_{1}$ and $\tau_{2}$ are $2.21 \pm 0.07ns$ and $19.25 \pm 0.68ns$ respectively. From this result, we conculde that the probe have a strong influence on the emission of parts of the QDs, while other QDs only changed a little.

As a comparation, a similar measurement was done for a fiber taper without AgNW. We pasted some QDs on the tip of a fiber taper and excited the QDs with objective lens. The fluorescence is detected from the output of the fiber. In this case, the experimental data (see Figure 3c) can be nicely fitted with $f_{1}$. Obtained lifetime is $19.02 \pm 0.26ns$, which is much larger than that with plasmonic probe collection. This is because the surface plasmon mode has a stronger field confinement, thus larger coupling strength with QDs.

\section{Influence of different mode}

In this section, we discuss in detail how the plasmonic modes influence the lifetime of the QDs. In previous works, people only theoretically studied the effect of the coupling between electric dipoles and different waveguide modes \cite{dipolemultimode}. The effect of each eigenmode in a multimode AgNW waveguide was experimentally investigated in our work. For a AgNW, there should exist three stationary modes, $TEM_{0}$, $TE_{1}$ and $TM_{1}$ (See Figure 1c). Since the later two higher-order modes are degenerate, here we take $TE_{1}$ as an illustration. The fundamental mode $TEM_{0}$ is strongly confined and symmetrically distributed near the nanowire surface. Its effective modes area scales as $r^{2}$, where $r$ is the radius of the nanowire. The higher-order modes have larger effective mode area compared to the fundamental mode. As shown in Figure 4a, the effective mode areas of the fundamental mode $TEM_{0}$ and the higher- order mode $TE_{1}$ are calculated. The $TE_{1}$ mode is always larger than the $TEM_{0}$ mode when $r<100\:nm$. Since the three modes have different electric field distributions, they couple with different QDs on the substrate. For $r<100\:nm$, the $TEM_{0}$ mode can only interact with QDs very close to the nanowire surface, while the $TE_{1}$ and $TM_{1}$ modes can interact with QDs in a scale of several hundred nanometers.

In our experiment, the probe is placed near the substrate. For QDs near the surface of the AgNW, it can interact with both $TEM_{0}$ and higher-order modes. As a result, we can detect a large decrease of the lifetime. For QDs far from the AgNW, it only interact with $TE_{1}$ and $TM_{1}$ modes with weak coupling strength, so the lifetime is relatively large. Our measured signal is a mixture of photons from QDs near and far from the AgNW surface. $A_{1}$ and $A_{2}$ are the weight of these two processes, which is relevant to the coupling strength, the number of QDs and the efficiency of the hybrid probe. In the experiment above, the ratio of $A_{1}$ to $A_{1}$ is $1.17$. Taken into consideration of the low transmittance of the fundamental mode, the value of $A_{1}/A_{2}$ should be larger. We also changed the distribution of QDs by pasting some QDs on the AgNW surface and moving the plasmonic probe away from the substrate. In this free standing case, all QDs are near the nanowire surface, and $A_{1}/A_{2}$ reached $3.93$, which proves that the fundamental $TEM_{0}$ mode dominates the coupling process.

Theoretically, the coupling strength $g$ between a two level quantum
emitter and an optical mode can be expressed as:
\begin{equation}
g \propto \sqrt{\frac{\hbar\omega}{2A_{eff}}\frac{|E(\overrightarrow{x})|}{max[E(\overrightarrow{x})]}}cos(\theta)
\end{equation}
where $A_{eff}$ is the effective mode area of the optical mode, $E(\overrightarrow{x})$ is the electric field of the optical mode, $\theta$ is the angle between the directions of the dipole and the field. According to Figure 4a, the electric field of $TE_{1}$ and $TM_{1}$ mode will diverge from the surface as the radius becomes smaller, while the field of $TEM_{0}$ mode will become more confined near the surface. Because the coupling strength has a negative correlation with the effective mode area, the coupling between the QD and the AgNW will increase significantly. Meanwhile, the collection area of the AgNW will become smaller, and this area is not restricted by the optical diffraction limit, which makes the local operation of quantum emitters possible.

\begin{figure*}[htb]
\includegraphics[width=14cm]{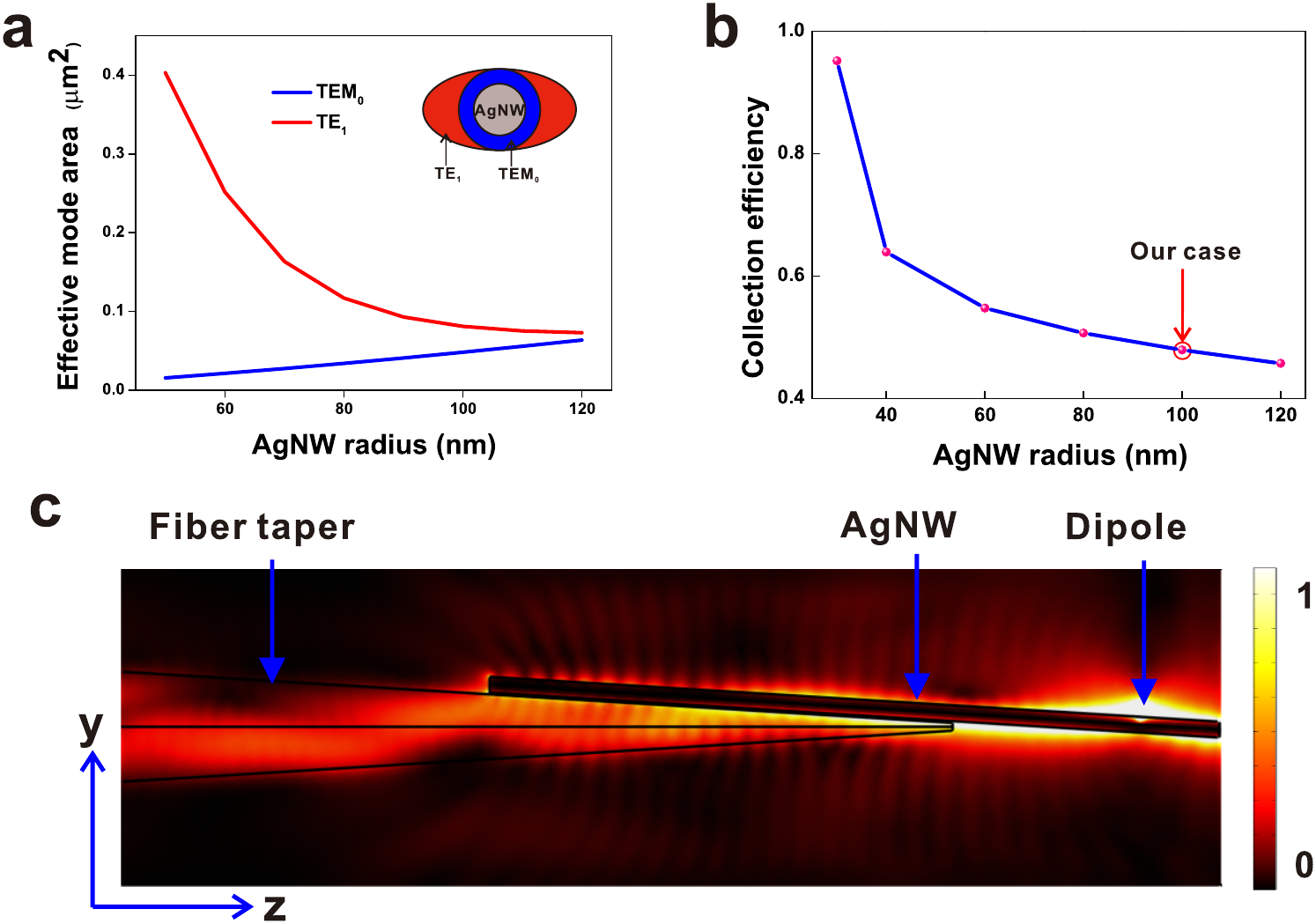}

\caption{(color online))Numerical simulation. (a) Effective mode area of the plasmonic modes. The inset is the schematic diagram of the plasmonic mode on AgNW. (b) Relation between the collection efficiency and the AgNW radius. The efficiency can even exceed $90\%$ when $r<30\:nm$. The efficiency is calculated by summing up of the energy flow along two directions of the AgNW. In fact, only about half of the energy can be guided to the fiber. (c) yz cut plane image of the 3D simulation. The radius of the nanowire is $100\:nm$. normal to the surface of the AgNW. }
\end{figure*}

\section{Efficiency of the plasmonic probe}

In the following, we will give a discussion about the efficiency of our plasmonic probe. There are two processes that determine the collection efficiency of the probe: the coupling between the QD and the AgNW, and the transmittance of light from the AgNW to the fiber taper. The efficiency of the first process mainly depends on the Purcell factor\cite{purcell} of the waveguide, while the the later process depends on the structure of the hybrid probe. Here we focus on the first process, and the calculation of the second process is given in the supporting information.

The QD is treated as an electric dipole and the coupling between an electric dipole and the plasmonic probe is simulated to characterize the efficiency of the probe. In our simulation, the dipole is placed $5\:nm$ away from the surface of the AgNW and the orientation is set normal to the surface. By integrating the total energy flow inside the simulation box and the energy flow along two directions of the AgNW, the collection efficiency of the $100\:nm$-radius AgNW is estimated to be $47.9\%$. Because the fluorescence propagates towards two directions, only half of the fluorescence is guided to the fiber taper. This energy loss can be avoided by putting the QD at the end of the probe, which results in a simulated efficiency of $56.3\%$ (see the supporting information). As the radius of the AgNW becomes smaller, the Purcell factor of the $TEM_{0}$ mode increases rapidly, thus the collection efficiency of the probe will increase significantly. When $r<30\:nm$, the collection efficiency of the AgNW can reach $90\%$. Figure 4b gives the relation between the collection efficiency and the nanowire radius.

Compared to the objective lens, which can only collect fluorescence emitted to one half space, the plasmonic probe can collect photons from the quantum emitters more efficiently. In our case, the total collection efficiency of the probe is $12.6\%$ according to the numerical simulation results. The relatively low efficiency in our experiment mainly originates from the absorption on the AgNW and the coupling between the AgNW and the fiber taper (see the supporting information). Both losses can be reduced by carefully designing the structure of the hybrid probe. In our simulation, we have neglected the influence of the silica substrate. The effective mode area of the $TEM_{0}$ mode will decrease significantly near a substrate \cite{AgnwmodeZou,AipingLiu}, thus the collection efficiency of the probe will be higher. Through the simulation and analysis above, we confirm that the probe can be a promising candidate to collect fluorescence of quantum emitters with high efficiency.

\section{Conclusions and Discussions}

In our experiment, we have realized the near-field collection of fluorescence from quantum emitters with a high-efficiency fiber-integrated plasmonic probe. The dramatic decrease of lifetime from $15.25ns$ to $2.21ns$ unambiguously demonstrated the strong near-field interaction between the QDs and the plasmonic modes. For the first time, we experimentally investigated the influence of each plasmonic mode on a multimode plasmonic waveguide. Our analysis on the lifetime of the fluorescence clarified the mechanisms of the coupling between the QDs and different surface plasmon modes. Theoretical simulation shows the high collection efficiency of the probe. The collection efficiency and the coupling strength can be further increased by reducing the diameter of the metallic nanowire or designing other structures with resonance \cite{goldnanotip}. An alternative way is to replace the nanowire by a metallic tapered tip \cite{DEchangprb}. By carefully designing the coupling between nanowire, and fiber and accurately controlling the position of the probe, we expect to realize highly efficient coupling between the plasmonic probe and single quantum emitters, which is promising to build a transistor in the single photon level \cite{singlephotontransistor}.

Due to the high confinement of the plasmonic mode, the probe only interacts with emitters localized near the nanowire, which can increase the signal to noise ratio. It is also promising to built a nanoscope and realize local optica operation of quantum emitters in nanometer scale. A possible application is to image luminous emitters beyond the diffraction limit. Meanwhile conversely, the probe can also be used to locally excite quantum emitters in nanoscale.

There are several benefits from the dielectric-metal nanoprobe: (1) The movable fiber-integrated plasmonic nano-probe gives a new way to collect photons from nano-emitters and can avoid complex combination of bulk optical components. (2) It is promising for near-field quantum probes, photonic endoscopes \cite{yangpeidong,livingcellendoscope} or movable single photon source. (3) The probe is broadband, with a very wide wavelength range. (4) The thermal property of silica fiber and AgNW also makes the probe's application at cryogenic temperature possible. At low temperature, the decrease of the loss of AgNW may even improve the performance of the probe.

In summary, the fiber-integrated plasmonic probe can both transport light forward and backward with high efficiency beyond the diffraction limit, and can be used to transport quantum states and realize strong light-matter interaction, which mays be useful in the area of nanophotonics and promising for quantum information devices. Our study encourages further investigations of quantum plasmonics and promotes the applications of surface plasmons in quantum optics field.

\subsection{Supporting Information}

The Supporting Information contains the fabrication method of the
probe, lifetime measurement process, simulated mode distribution of
AgNW and fiber taper, simulation results of the backward transport
of light from nanowire to fiber taper and the simulation result of
the collection efficiency of the AgNW when QD is placed at the end
of the wire. This material is available free of charge via the Internet
at http://pubs.acs.org.

\begin{acknowledgement}

This work was funded by the National Key R \& D Program (Grant No.2016YFA0301700), Innovation Funds from the Chinese Academy of Sciences (grant no. 60921091); NNSFC (grant no. 11374289, 61590932, 61505195); the Fundamental Research Funds for the Central Universities, the Open Fund of the State Key Laboratory on Integrated Optoelectronics (IOSKL2015KF12). The authors thank C. H. Dong and F. W. Sun for technical support and useful discussion.

\end{acknowledgement}

\section{Notes}

The authors declare no competing financial interests.

\end{document}